\documentclass[%
 aps, pra, reprint,
 superscriptaddress, 
 showpacs, nopreprintnumbers,
 longbibliography,
 amsmath, amssymb, floatfix
]{revtex4-2}

\usepackage{graphicx}
\usepackage{dcolumn}
\usepackage{bm}
\usepackage{hyperref}
\usepackage{physics}
\usepackage[usenames,dvipsnames]{xcolor}
\usepackage{booktabs}

\newcommand{\beginsupplement}{%
        \setcounter{table}{0}
        \renewcommand{\thetable}{S\arabic{table}}%
        \setcounter{figure}{0}
        \renewcommand{\thefigure}{S\arabic{figure}}%
     }

\newcommand{\mytitle}{Dipolar Bose-Hubbard model in finite-size real-space cylindrical lattices}

\begin{document}

\title{\mytitle}

 \author{Michael Hughes}
  \email{michael.hughes@physics.ox.ac.uk}
  \affiliation{Clarendon Laboratory, University of Oxford, Parks Rd, Oxford OX1 3PU, United Kingdom.}

 \author{Dieter Jaksch} 
 \affiliation{Clarendon Laboratory, University of Oxford, Parks Rd, Oxford OX1 3PU, United Kingdom.}
 \affiliation{Institut f\"{u}r Laserphysik, Universit\"{a}t Hamburg, 22761 Hamburg, Germany.}

\date{\today}

\begin{abstract}
Recent experimental progress in magnetic atoms and polar molecules has created the prospect of simulating dipolar Hubbard models with off-site interactions. When applied to real-space cylindrical optical lattices, these anisotropic dipole-dipole interactions acquire a tunable spatially-dependent component while they remain translationally-invariant in the axial direction, creating a sublattice structure in the azimuthal direction. We numerically study how the coexistence of these classes of interactions affects the ground state of hardcore dipolar bosons at half-filling in a finite-size cylindrical optical lattice with octagonal rings.
When these two interaction classes cooperate, we find a solid state where the density order is determined by the azimuthal sublattice structure and builds smoothly as the interaction strength increases. For dipole polarisations where the axial interactions are sufficiently repulsive, the repulsion competes with the sublattice structure, significantly increasing entanglement and creating two distinct ordered density patterns. The spatially-varying interactions cause the emergence of these ordered states in small lattices as a function of interaction strength to be staggered according to the azimuthal sublattices.

\end{abstract}
\maketitle

\section{Introduction}

The Bose-Hubbard model has generated significant interest as a simple model to produce a quantum phase transition between Mott insulator and superfluid states which can be simulated by ultracold atoms with short-range interactions in optical lattices \cite{boson_transition,dieter_cold,SF_to_MI_experiment}. The development of optically-trapped systems with dipolar interactions, such as magnetic atoms \cite{magnetic_atoms_lattice,EBH_atoms,magnetic_atoms_large_spin}, polar molecules \cite{polar_molecules_lattice_3,  polar_molecules_lattice,molecules_dense, polar_molecules_Durham, polar_molecules_lattice_2,polar_molecules_lattice_4,polar_molecules_frontiers}, and Rydberg atoms \cite{rydberg_lattice,recent_rydberg,recent_rydberg_2} has drawn attention to more complex models. In particular, dipolar bosons with long-ranged and anisotropic dipole-dipole interactions can additionally simulate extended Bose-Hubbard models featuring off-site interactions \cite{EBH_atoms}, which are predicted to stabilise density wave and supersolid phases \cite{extended_ss,extended_ss_2,extended_ss_3,supersolid_square,supersolid_triangle_neighbour,supersolid_triangle,novel_phases,extended_BH_tutorial,extended_BH_review}. 
The production of curved trap geometries, such as ring traps and lattices \cite{ring_trap_BEC, experimental_ring_trap_1, experimental_ring_trap_2, experimental_ring_trap_3, ring_lattice}, combined with the anisotropy of the dipole-dipole interaction also allows the possibility of spatially-varying interactions, as studied for gases in ring, torus, and spherical-shell potentials \cite{dipolar_fermions_ring,gas_ring,condensate_torus,condensate_ring_or_shell, dipolar_BEC_shell, dipolar_BEC_shell_2} and for bosons in a discrete octagonal ring lattice \cite{ring}, which leads to density modulation imposed directly by the interaction variation.

 Following proposals for short real-space cylindrical optical lattices \cite{cylindrical}, we numerically investigate a finite-size extended Bose-Hubbard model which uses the cylindrical lattice geometry to create spatially-varying interactions. The addition of the translationally-invariant axial interactions between rings causes tunable cooperation and competition between spontaneous ordering and sublattice-imposed ordering. Modifying the dipole polarisation direction changes the magnitude of the spatial variation of interactions in the azimuthal direction while tuning the sign and magnitude of the axial interactions. For attractive axial interactions, the ground state behaviour is determined by the strength of the sublattice ordering in the azimuthal direction in creating its own density wave pattern, which is qualitatively similar to the single octagonal ring. The cylindrical geometry demonstrates richer physics when the dipole polarisation is angled far from the cylinder axis and repulsive axial interactions frustrate the sublattice structure, leading to greater entanglement and importance of next-nearest-neighbour interactions.

This paper is arranged in the following manner: In section \ref{Hamsec}, we introduce the lattice Hamiltonian for a hard-core cylindrical dipolar Bose-Hubbard model and describe our numerical calculations and order parameters for the specific case of octagonal rings. In section \ref{ressec}, we present our numerical results for the finite-size `phase diagram' of the octagonal-prism system and describe the ground states. In section \ref{ressec_fss} we study three different transitions from section \ref{ressec} at varying cylinder lengths. In section \ref{discsec} we discuss extensions to this model and comment on its possible physical realisation.   

\section{Hamiltonian}
\label{Hamsec}

\begin{figure}
\includegraphics[scale=0.25]{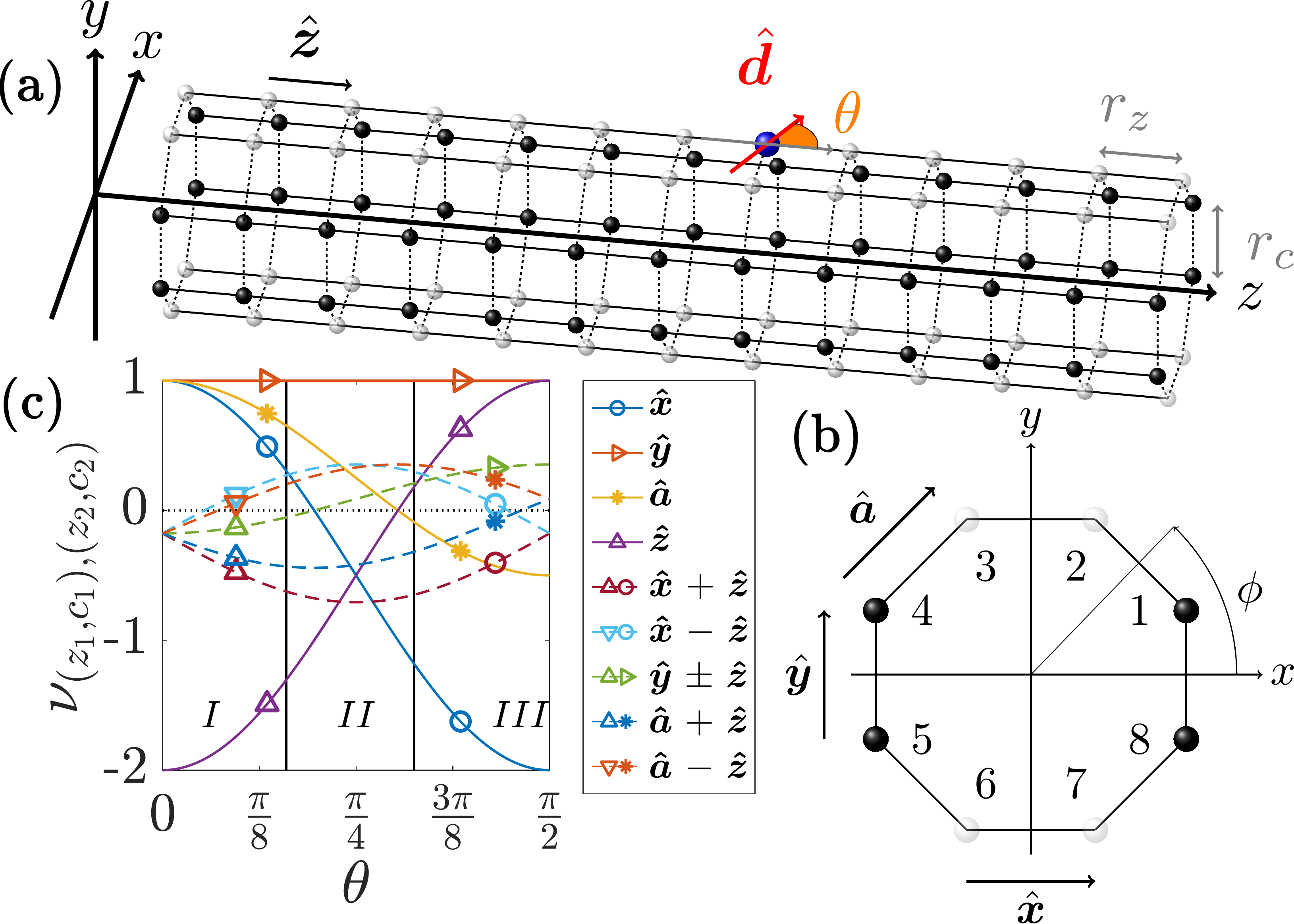}
\caption{Colour online. (a) Cylindrical lattice formed by $13$ octagonal rings giving the parameters $L_z = 13$ and $L_c = 8$. The separations between adjacent sites along the axis and around the ring are denoted by $r_z$ and $r_c$ respectively. A single boson, represented by the larger sphere, occupies one of the lattice sites. Its dipole moment $\hat{\bm{d}}$, shown by the single-headed arrow, is polarised in the $x-z$ plane at an angle $\theta$ to the $z$ axis. (b) A single ring of the lattice, showing azimuthal angle $\phi$ and $L_c = 8$ sites labelled by their $c$ coordinate. The two sublattices (named polar and equatorial) are denoted by white and black spheres respectively. The different nearest-neighbour in-plane separation vectors are labelled. (c) Relative dipole-dipole interaction strengths between nearest-neighbour (solid lines) and next-nearest-neighbour (dashed lines) sites as a function of polarisation angle $\theta$. Each interaction has two markers as a visual aid. Domain is split by black vertical lines into three regions ($I$,$II$,$III$) according to the qualitative ordering mechanism and ground state at strong interaction. Zero interaction strength is shown by the black dotted line.}
\label{diagramfig}
\end{figure}

A schematic of a polarised boson in the cylindrical lattice is shown in Figure \ref{diagramfig}(a). The cylindrical lattice sites are specified by their axial index $z$ and azimuthal index $c$. In this model, polarised bosons are able to hop between adjacent sites in two directions: along the cylinder axis, denoted by the black lattice lines in Figure \ref{diagramfig}(a), or around the ring of the cylinder, denoted by the dashed lattice lines. The bosons are all polarised in the same direction and interact due to the dipole-dipole interaction. We consider the hard-core limit, in which the bosons have such strong on-site repulsion that there can only be zero or one particles per site. This means the Hamiltonian is
\begin{equation}
\begin{split}
H = & -J \sum_{\langle (z_1,c_1),(z_2,c_2) \rangle} \hat{b}^{\dagger}_{z_1,c_1} \hat{b}_{z_2,c_2} + H.c. \\
&+ \frac{V}{2}\sum_{(z_1,c_1) \neq (z_2,c_2)} \nu_{(z_1,c_1),(z_2,c_2)} \hat{n}_{z_1,c_1} \hat{n}_{z_2,c_2}, 
\end{split}
\label{Heq}
\end{equation}
where $H.c$ means Hermitian conjugate, $z_k,c_k$ denote the axial and azimuthal coordinates of site $k$, $\langle (z_1,c_1),(z_2,c_2) \rangle$ are neighbouring sites, $\hat{b}$ is the hard-core boson annihilation operator such that $\hat{b}^2 = (\hat{b}^{\dagger})^2 = 0$, $\hat{n} = \hat{b}^{\dagger}\hat{b}$ is the on-site density, $J$ is the tunnelling amplitude, $V$ is the overall dipole-dipole interaction strength, and $\nu_{(z_1,c_1),(z_2,c_2)}$ is a factor describing the distance decay and orientation dependence of the interaction which is given by
$\nu_{(z_1,c_1),(z_2,c_2)} = (1-3\cos^2(\alpha_{(z_1,c_1),(z_2,c_2)}))/r_{(z_1,c_1),(z_2,c_2)}^3 $,
where $r_{(z_1,c_1),(z_2,c_2)}$ is the distance between two sites, and $\alpha_{(z_1,c_1),(z_2,c_2)}$ is the angle between the common dipole polarisation direction and the separation vector of these sites. We have assumed that the tunnelling amplitude $J$ is equal around the cylinder and along the axis, and we express all energies in units of $J$ throughout the rest of the paper, which is equivalent to setting $J = 1$ in equation \ref{Heq}.

The important property of the curved surface of the cylindrical optical lattice is that the 3D separation vector between sites with a given displacement along the 2D lattice surface depends on the azimuthal coordinate $c$. Due to the angular dependence of the dipole-dipole interaction, this means the DDI changes magnitude and sign around cylinder ring.

\subsection{Numerical Calculations}

While this Hamiltonian offers a wide class of models, we focus on a specific realisation for concreteness in our numerical calculations. We assume that the lattice spacing between adjacent sites on around the ring of the cylinder is equal to the spacing along the axis ($r_c = r_z = 1$). We consider a cylinder made of octagonal rings ($L_c = 8$), which is aligned along the $z$ direction and oriented with respect to the in-plane $x$ and $y$ axes as shown in Figure \ref{diagramfig}(b). We note that for fixed $L_c$, the thermodynamic limit is $L_z \rightarrow \infty$ and is therefore one-dimensional.

We constrain the dipole polarisation to the $x-z$ plane and parametrise it using the angle $0 \leq \theta \leq \frac{\pi}{2}$, where $\theta = 0$ corresponds to polarisation in the $z$ direction and $\theta = \frac{\pi}{2}$ corresponds to the $x$ direction. The strengths of the nearest-neighbour and next-nearest-neighbour dipole-dipole interactions are shown as a function of $\theta$ in Figure \ref{diagramfig}(c). There are four different nearest-neighbour and five different next-nearest-neighbour interactions due to the curvature of the lattice in the $x-y$ plane, compared to a square lattice which has at most two different nearest-neighbour and two different next-nearest-neighbour interactions for a given polarisation angle. With polarisation confined to the $x-z$ plane, the spatial variation of interactions when $\theta \neq 0$ creates two sublattices of inequivalent sites on each ring which we label `polar' and `equatorial' as denoted in Figure \ref{diagramfig}(a) and (b). Both classes of sites connect to lattice vectors in the $\hat{\bm{a}}$ and $\hat{\bm{z}}$ directions but the polar sites additionally connect to lattice vectors in the $\hat{\bm{x}}$ direction while the equatorial sites connect to lattice vectors in the $\hat{\bm{y}}$ direction, meaning the equatorial sites experience more repulsive dipolar interactions. While the dipole-dipole interaction in equation \ref{Heq} in principle causes interactions between all pairs of sites, we include only the interactions up to next-nearest-neighbour shown in Figure \ref{diagramfig}(c) for numerical reasons, noting that the interactions decay quickly with distance. 

For numerical reasons and to investigate the effect of the cylinder edges, we first focus on the ground state behaviour as a function of interaction strength and polarisation direction for systems with a small number of sites in the axial direction $(10 \leq L_z \leq 16)$ in section \ref{ressec}. Unless stated, we present results for $L_z = 13$ because we find that certain ordered states are less disrupted by the cylinder edges. For simplicity, we perform our calculations with fixed particle number $N$, which is set to half the number of sites $N = N_H = \frac{L_z L_c}{2}$ unless stated. In section \ref{ressec_fss} we present results for varying $L_z$ with the same fixed filling fraction.

We use DMRG to approximate the ground state of this system \cite{DMRG_White,DMRG_White_2} using bond dimensions $\chi \leq 6400$, implementing the numerics in the ITensor library \cite{itensor}. We include further numerical details in appendix section \ref{calcsec}. 
\subsection{Observables}

To categorise the ground states we extract physical quantities from our numerical results. Firstly, to investigate density wave states, we use density-density correlations $\langle \hat{n}_{z = z_1,c = c_1} \hat{n}_{z = z_1 + \Delta_z,c = c_1 + \Delta_c} \rangle$ where we map $c$ to values $1 \leq c \leq L_c$ using the periodic boundary conditions if $c_1 + \Delta_c > L_c$. For a given displacement $\{\Delta_z,\Delta_c\}$, we take the average density-density correlation for all such pairs of sites in the `bulk', which we define by removing three rings from both edges, and normalise using the formula

\begin{equation}
M_{\Delta_z,\Delta_c} = \frac{\sum_{\{z_1,c_1\}} \langle \hat{n}_{z_1,c_1}\hat{n}_{z_1+\Delta_z,c_1+\Delta_c}\rangle - n_b n_0^2}{n_b n_0^2} \label{nncorreq},
\end{equation}
where $\{z_1,c_1\}$ is a site where both $(z_1,c_1)$ and $(z_1 + \Delta_z, c_1 +\Delta_c)$ are in the bulk, $n_b$ is the number of such pairs of sites, and $n_0$ is the average density on the included sites $(z_1,c_1)$. This means that $M_{\Delta_z,\Delta_c}$ takes the value $1(-1)$ if the presence of a particle on site $(z_1,c_1)$ implies the presence(absence) of a particle on site $(z_1 + \Delta_z, c_1 + \Delta_c)$, while it is zero if there is no correlation. We also define sublattice-resolved versions of the $M_{0,4}$ parameter, where $M_{0,4,P}$($M_{0,4,E}$) is calculated using only pairs of sites on the polar(equatorial) sublattice. 

Additionally, we look for a large population fraction in the dominant bosonic orbital and inter-site correlations which decay polynomially with distance in the $\hat{z}$ direction as signatures of a superfluidity.  These quantities are accessible numerically through the eigenvalues and elements of the one-body density-matrix $\langle \hat{b}^{\dagger}_{z_1,c_1} \hat{b}_{z_2,c_2} \rangle$  respectively. While the population fraction in the dominant bosonic orbital vanishes in the one-dimensional thermodynamic limit for hard-core bosons, its slower decay with system length in the superfluid state (even in a strictly one-dimensional chain \cite{hard_core_boson_1D_condensation_1, hard_core_boson_1D_condensation_2}) than in the solid state ($\propto L_z^{-1}$) offers contrast in finite-length systems. For a clearer distinction, we define the difference $\delta_e$ between the population fractions in the largest two modes, which would give $\delta_e = 1$ in a perfect condensate and $\delta_e = 0$ in a solid \cite{checker}. Again, we define sublattice-resolved versions of this quantity, where $\delta_{e,E}$($\delta_{e,P}$) denotes the difference in population fraction on the polar(equatorial) sublattice between the two modes with the greatest population on that sublattice.  

A useful quantity accessible to MPS is the Von-Neumann entanglement entropy across a lattice bipartition. This is calculated using the formula
\begin{equation}
S_{ent} = -\sum_{\nu = 1}^{\chi} |\lambda_{\nu}|^2 \ln(|\lambda_{\nu}|^2), \label{senteq}
\end{equation}
where $\lambda_{\nu}$ are the coefficients of the Schmidt decomposition (of which at most the largest $\chi$ are kept in the MPS). Peaks in entanglement entropy can be used to pinpoint transitions between superfluid and density wave states without choosing the order parameters for the phases in advance \cite{checker,DMRG_rydberg_square,DMRG_rydberg_kagome}. The first derivative of entanglement entropy (and other entanglement measures) has also been used to locate transitions which do not feature discontinuities in order parameters \cite{entanglement_transition_experiment,entanglement_entropy_XXZ,entanglement_review}. The bipartition we use cuts the lattice across the axis into subsystems of as equal length as possible, which for $L_z = 13$ means subsystems of six and seven rings respectively (i.e. one subsystem contains all sites with $1 \leq z \leq 6$). While observing the Von-Neumann entanglement entropy is experimentally challenging, we note that the second-order Renyi entanglement entropy (which provides a lower bound for $S_{ent}$) has been measured in Bose-Hubbard systems in optical lattices \cite{experimental_Renyi_entropy}.

\section{Ground State Diagram}
\label{ressec}

\begin{figure}
\includegraphics[scale=0.6]{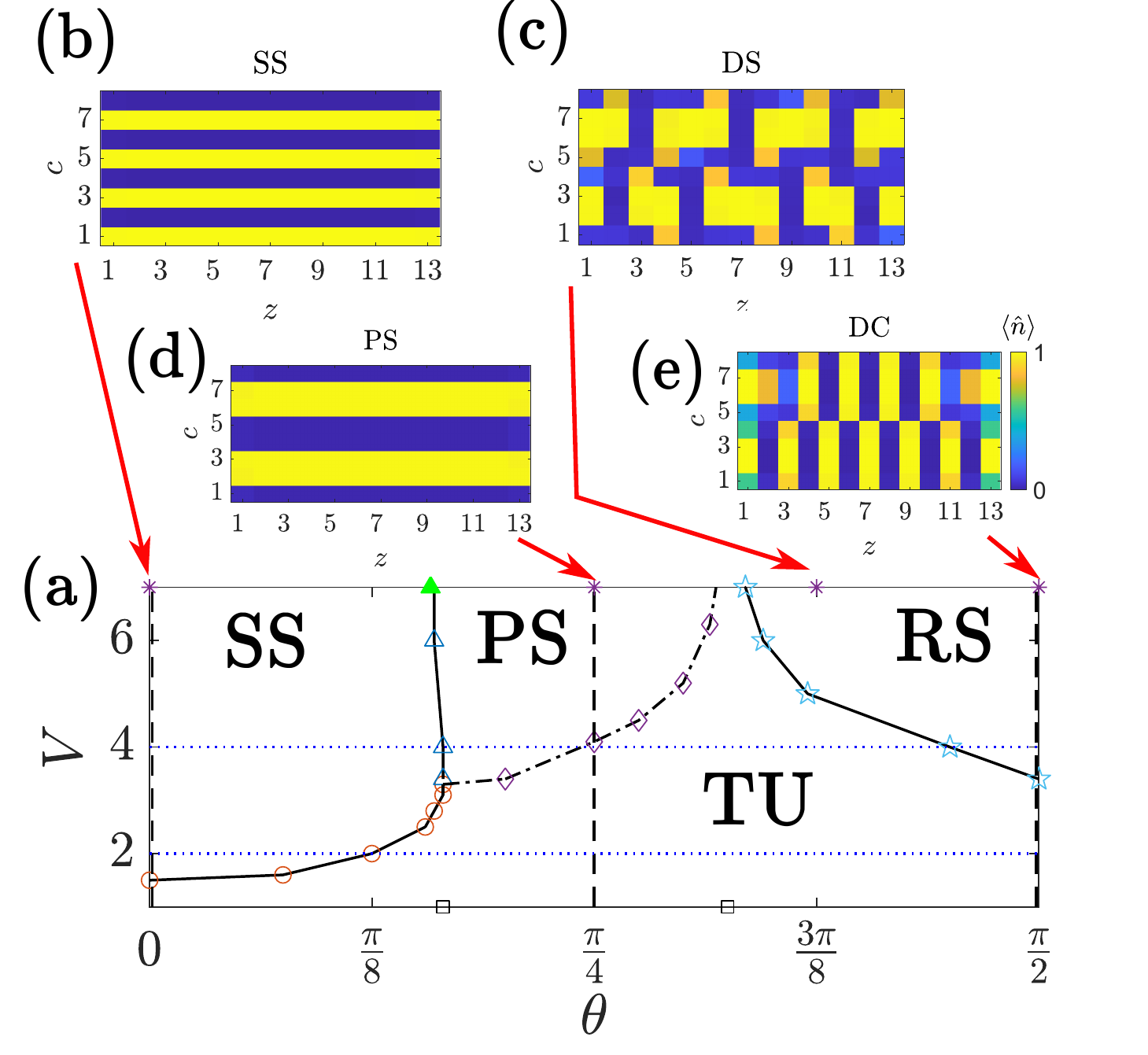}
\caption{Colour online. (a) Finite-size phase diagram of model. Unfilled markers show calculated finite-size transition points, while black lines join neighbouring markers as a guide to the eye (the boundary between the PS and TU regions is marked in a dash-dotted line to indicate that the transition is not sharp). Filled triangle shows the value of $\theta$ for transition from SS to PS as $V \rightarrow \infty$. Quantitative order parameters are plotted for the dashed lines at $\theta = \{0, \frac{\pi}{4}, \frac{\pi}{2}\}$ in Figure \ref{Vfig} and for the dotted lines at $V = \{2, 4\}$ in Figure \ref{thetafig}. Black square markers on the $\theta$ axis mark the boundaries between regions $I$, $II$, and $III$.
(b-e) Expectation value of on-site density for solid ordered states using the common colour scale shown next to (e). Arrows point towards the corresponding parameters in (a) marked by asterisks.}
\label{phasefig}
\end{figure}

\begin{figure}
\includegraphics[scale=0.5]{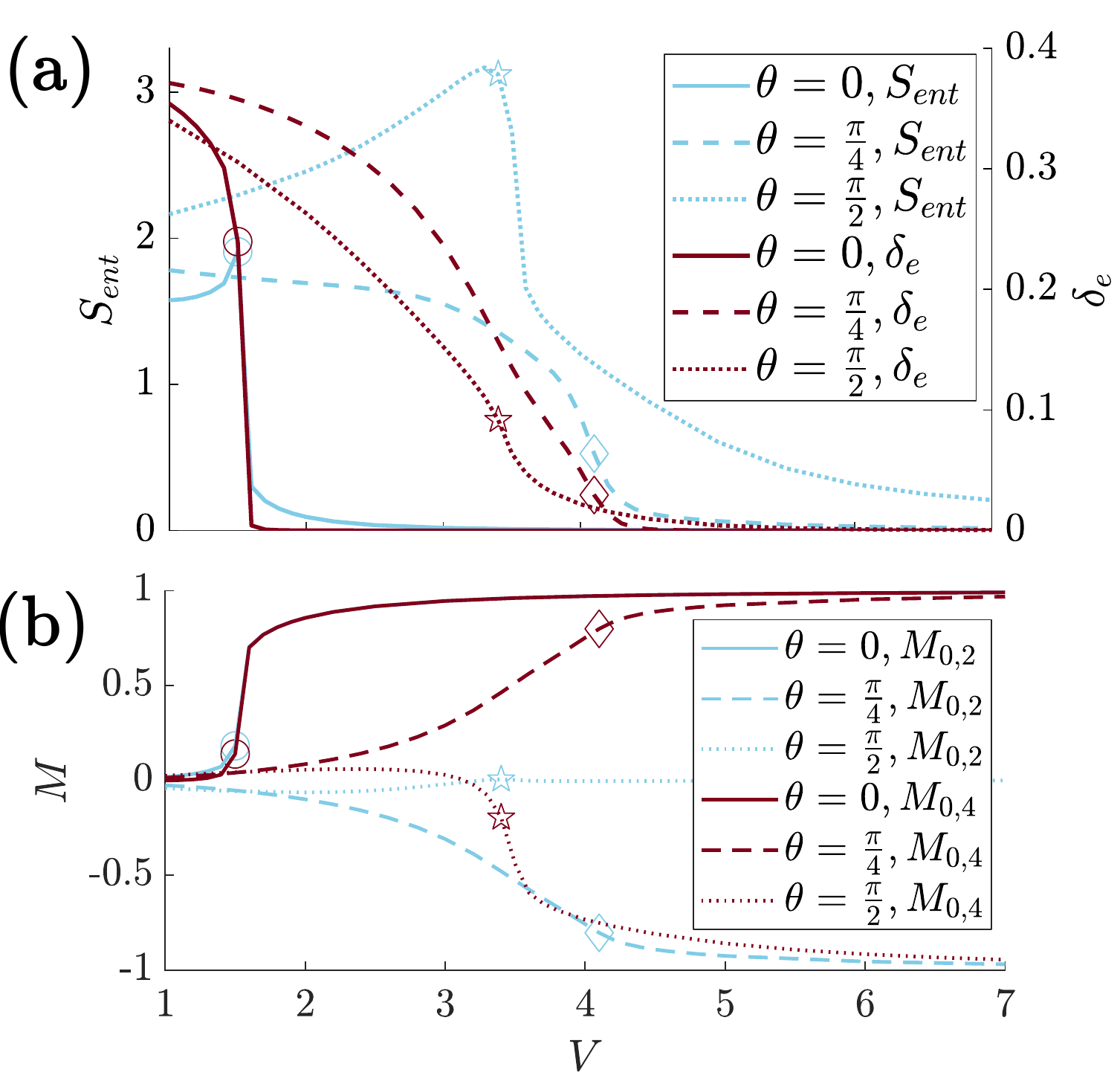}
\caption{Colour online. Physical parameters for $\theta = 0$, $\frac{\pi}{4}$, and $\frac{\pi}{2}$ as a function of $V$ corresponding to the black dashed lines in Figure \ref{phasefig}(a). (a) $S_{ent}$ on the left-hand y-axis and $\delta_e$ on the right-hand y-axis. (b) Solid order parameters $M_{\Delta_z,\Delta_c}$. Any transition lines from Figure \ref{phasefig}(a) which are intersected by these graphs are denoted with the corresponding marker.}
\label{Vfig}
\end{figure}

\begin{figure}
\includegraphics[scale=0.5]{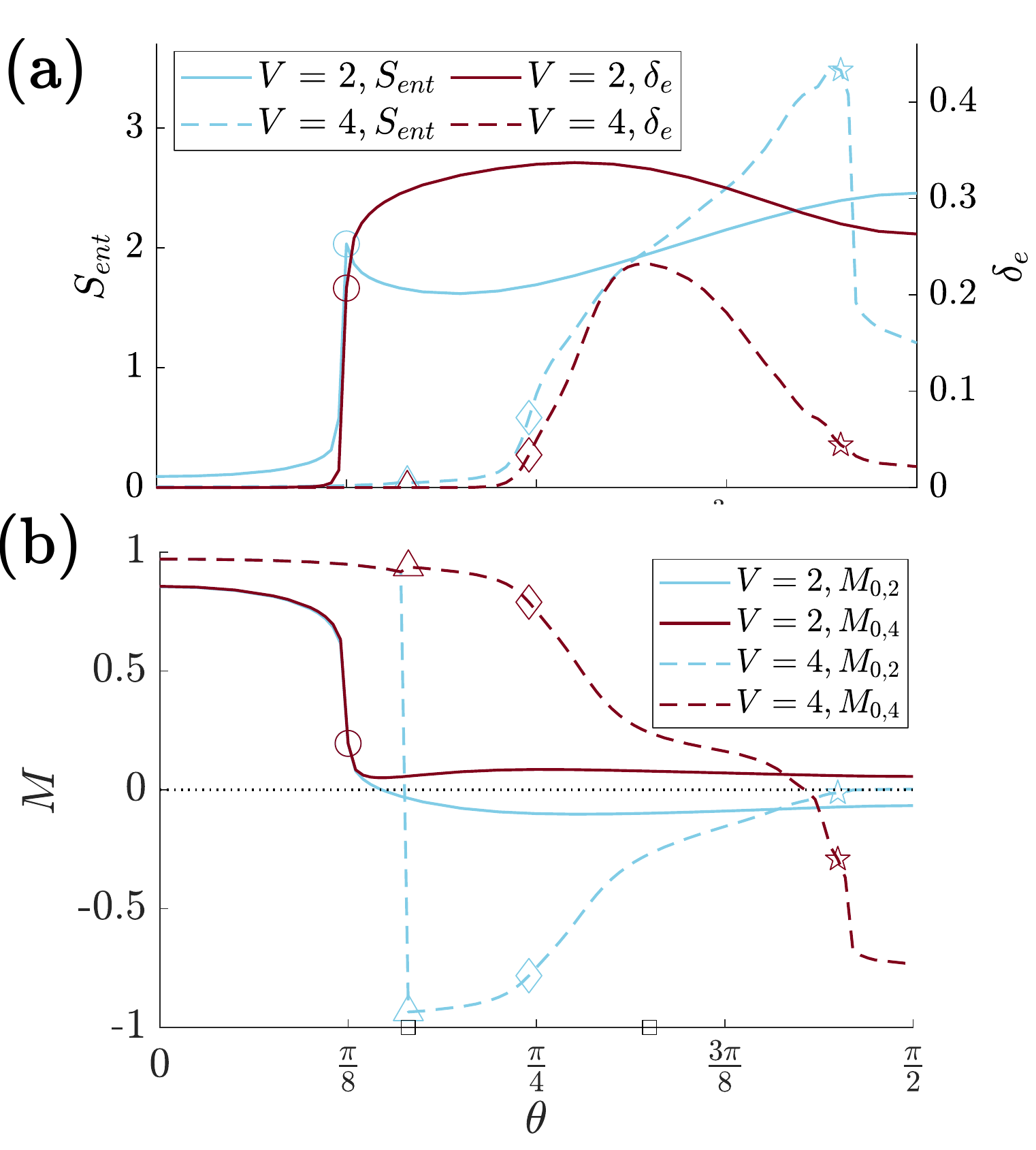}
\caption{Colour online. Physical parameters for $V = 2$ and $V = 4$ as a function of $\theta$ corresponding to the dotted lines in Figure \ref{phasefig}(a). (a) $S_{ent}$ on the left-hand y-axis and $\delta_e$ on the right-hand y-axis. (b) Solid order parameters $M_{\Delta_z,\Delta_c}$. A dotted black line at $M_{\Delta_z,\Delta_c} = 0$ is added as a guide to the eye. Any transition lines from Figure \ref{phasefig}(a) which are intersected by these graphs are denoted with the corresponding marker. The boundaries between regions $I$, $II$, and $III$ are denoted by black square markers on the $\theta$ axis.}
\label{thetafig} 
\end{figure}

This model supports a number of qualitatively different ground states as a function of the interaction strength $V$ and polarisation direction $\theta$, which are shown in Figure \ref{phasefig}. We increment $V$ in steps of $0.1$ and $\theta$ in steps of $0.01 \frac{\pi}{2}$ to identify all features in Figure \ref{phasefig}(a). The quantitative behaviour of the observables along a number of cuts across this diagram is shown in Figures \ref{Vfig} and \ref{thetafig}. To summarise the states which we quantitatively analyse, we show the expected values of the observables in Table \ref{ordtab}, although we emphasise again that $\delta_e$ is expected to vanish slowly in the one-dimensional thermodynamic limit. The physics of each state and the processes to locate finite-size phase boundaries are described in this section.

\begin{table}
\begin{tabular}{lrrrrr}
\hline
   & $\delta_e$          & $S_{ent}$            & $M_{\Delta_z = 0,\Delta_c = 2}$ & $M_{\Delta_z = 0,\Delta_c = 4}$ \\
\hline 
TU & $>0$ & $>0$ & $\approx 0$   & $\approx 0$   \\
\hline
SS & $0$               & $0$               & $+1$   & $+1$  \\
\hline
PS & $0$               & $0$               & $-1$  & $+1$    \\
\hline
DC & $0$               & $0$               & $0$   & $-1$  
	
\end{tabular}
\caption{Expected values of physical parameters in quantitatively analysed states for finite-length systems.}
\label{ordtab}
\end{table}

For weak dipole-dipole interaction, the tunnelling energy is the most important, leading to a tunnelling-dominated (TU) ground state with a large population fraction in a single bosonic orbital and weak density modulations. At $\theta = 0$, this state demonstrates polynomial decay of correlations which indicates superfluidity, while this quantitative relationship is gradually lost at higher $\theta$ as small density modulations and greater entanglement emerge. For strong dipole-dipole interaction, the particles are fixed in position and the ground state forms different solid orderings depending on $\theta$, which are favoured in the regions denoted $I$, $II$, and $III$ in Figure \ref{diagramfig}(c) respectively.

\subsection{Region $I$}

For low $\theta$ (region $I$), the spatial dependence of interactions is weak (non-existent at $\theta = 0$) and the physics is similar to a a flat square lattice Bose-Hubbard model with dipole polarisation along one of the lattice vectors \cite{QMCstripe,in_plane_BH_noSS}. The dipole-dipole interaction is attractive between rings and repulsive within rings, meaning that the particles line up along the $z$ axis in a `stripe solid' (SS) state at strong interaction (Figure \ref{phasefig}(b)). In the one-dimensional thermodynamic limit, the only symmetry broken by the TU-SS transition is a discrete sublattice symmetry, meaning the transition is expected to be second-order. 

This transition is characterised by the dramatic increase in solid order and decrease in $\delta_e$ as a function of $V$, as seen in Figure \ref{Vfig}(a) for $\theta = 0$. This coincides with the peak in $S_{ent}$ that we use to determine the boundary which is marked with circles in Figure \ref{phasefig}(a). $S_{ent}$ takes a low value on the solid side of the phase transition because DMRG favours the two product states in the two-fold degenerate ground state subspace over their entangled superpositions \cite{2D_DMRG_review} while the TU has significant entanglement entropy because it has local density fluctuations while the global particle number is conserved \cite{global_symmetry_entanglement}.

\subsection{Region $II$}

Moving into region $II$, the attractive sum of the next-nearest-neighbour interactions within the `polar' sublattice overcome the repulsive interactions in the $\vec{x}$ direction, causing the particles to occupy only the polar sublattice. Due to edge effects, the polar sites at $c = 2,7$ on the first ring ($z = 1$) and $c = 3,6$ on the final ring ($z = L_z$) are most vulnerable to reduced occupation due to finite tunnelling out of all nominally-occupied sites. Unlike the stripe solid which occupies one of two degenerate sublattices spontaneously, the `polar stripe' (PS) (Figure \ref{phasefig}(c)) occupies the single sublattice which has lower energy due to the spatially-varying Hamiltonian. This is similar to the density-wave states seen in Bose-Hubbard models with alternating potentials on a square lattice \cite{alternating_square_BH3,alternating_square_BH1,alternating_square_BH2}, although here the sublattice structure is imposed by the dipolar interactions rather than an external potential. 

As a result, the TU-PS transition does not break a discrete sublattice symmetry and has gentler changes in order parameters than the TU-SS transition as shown in the $\theta = \frac{\pi}{4}$ lines in Figure \ref{Vfig} and the $V = 4$ lines in Figure \ref{thetafig}, similar to the transition between a Mott insulator and a superfluid in extended Bose-Hubbard models on square lattices \cite{extended_repBH_softcore}. Unlike the other solid-tunnelling transitions in this model, it is not accompanied by a peak in $S_{ent}$, but still features an increase in $S_{ent}$ and the occupation of the lowest bosonic orbital and a decrease in the magnitude of the density-density correlation order parameters.

While the transition is very smooth for small systems, we determine the nominal boundary in Figure \ref{phasefig}(a) by locating the sharpest changes in the order parameters and wavefunction. Specifically, for fixed $\theta$ we place a diamond marker at the value of $V$ for which $-\frac{\partial S_{ent}}{\partial V}$, calculated using the central difference method, reaches a local maximum. These markers coincide (within our minimum $V$ increment of $0.1$) with a local minimum in the absolute value of the inner product between equally-spaced neighbouring ground state wavefunctions, indicating a local maximum of the fidelity susceptibility, which has also been used to pinpoint transitions \cite{DMRG_rydberg_kagome,fidelity_transitions_general,fidelity_transitions_review,fidelity_transition_MPS}. A more precise description of this transition is enabled by the larger cylinder lengths studied in section \ref{resssec_TUPS}.

The transition from the SS to the PS is found through an abrupt change as a function of $\theta$ in the density-density correlation order parameter $M_{\Delta_z = 0,\Delta_c = 2}$ as defined in equation \ref{nncorreq}. This identifies the transition because in the SS(PS), the presence of a particle on a given site implies the presence(absence) of a particle two sites further around the ring of the cylinder. The $M_{\Delta_z = 0,\Delta_c = 4}$ order parameter is unaffected by this transition, as are $S_{ent}$ and $\delta_e$. This is shown in the $V = 4$ lines in Figure \ref{thetafig}.

For a cylinder of infinite length in the axial direction and DDI truncated to next-nearest-neighbour, this transition should happen at $\theta \approx 0.307 \frac{\pi}{2}$ when $V >> J$, while for $L_z = 13$, edge effects favour the SS and push the transition to $\theta \approx 0.314 \frac{\pi}{2}$ (calculation details in appendix section \ref{sspssec}), which is consistent with our numerical results for $V = 6$ which show the SS at $\theta = 0.31 \frac{\pi}{2}$ and the PS at $\theta = 0.32 \frac{\pi}{2}$ . For lower values of $V$, this transition can happen at a slightly larger value of $0.32 \frac{\pi}{2} \leq \theta \leq 0.33 \frac{\pi}{2}$ as the SS is favoured by tunnelling at the lattice edges. The lowest value of $\theta$ for which the PS (rather than the SS) is found is recorded as the transition point and is denoted by the triangular markers in Figure \ref{phasefig}(a).

\subsection{Region $III$}

Moving into region $III$ at $\theta \geq 0.65 \frac{\pi}{2}$, there are attractive interactions in the $\hat{\bm{x}}$ and $\hat{\bm{a}}$ directions while the interactions between rings are repulsive. Although the attractive interactions within the rings favour the polar sublattice even more strongly than in region $II$, the repulsive interactions along the axis repel particles to the cylinder edges and penalise occupying the polar sublattice on neighbouring rings.

At weak interaction, the TU state maintains its weak density modulations in the bulk but smoothly acquires significantly increased $S_{ent}$ as shown in Figure \ref{thetafig} (a), as the frustration caused by the axial repulsion induces large density-density correlations, particularly for the polar sites. At strong interaction, region $III$ features states where occupation of the dominant bosonic orbital is suppressed, even in small systems, by dipolar interactions but which are often difficult to categorise in a finite-size system due to edge effects preventing density order. We denote this set of ground states as a `repulsive solid' (RS) after the repulsive axial interactions and now describe the states within it which do demonstrate regular order.

\begin{figure}
\includegraphics[scale=0.55]{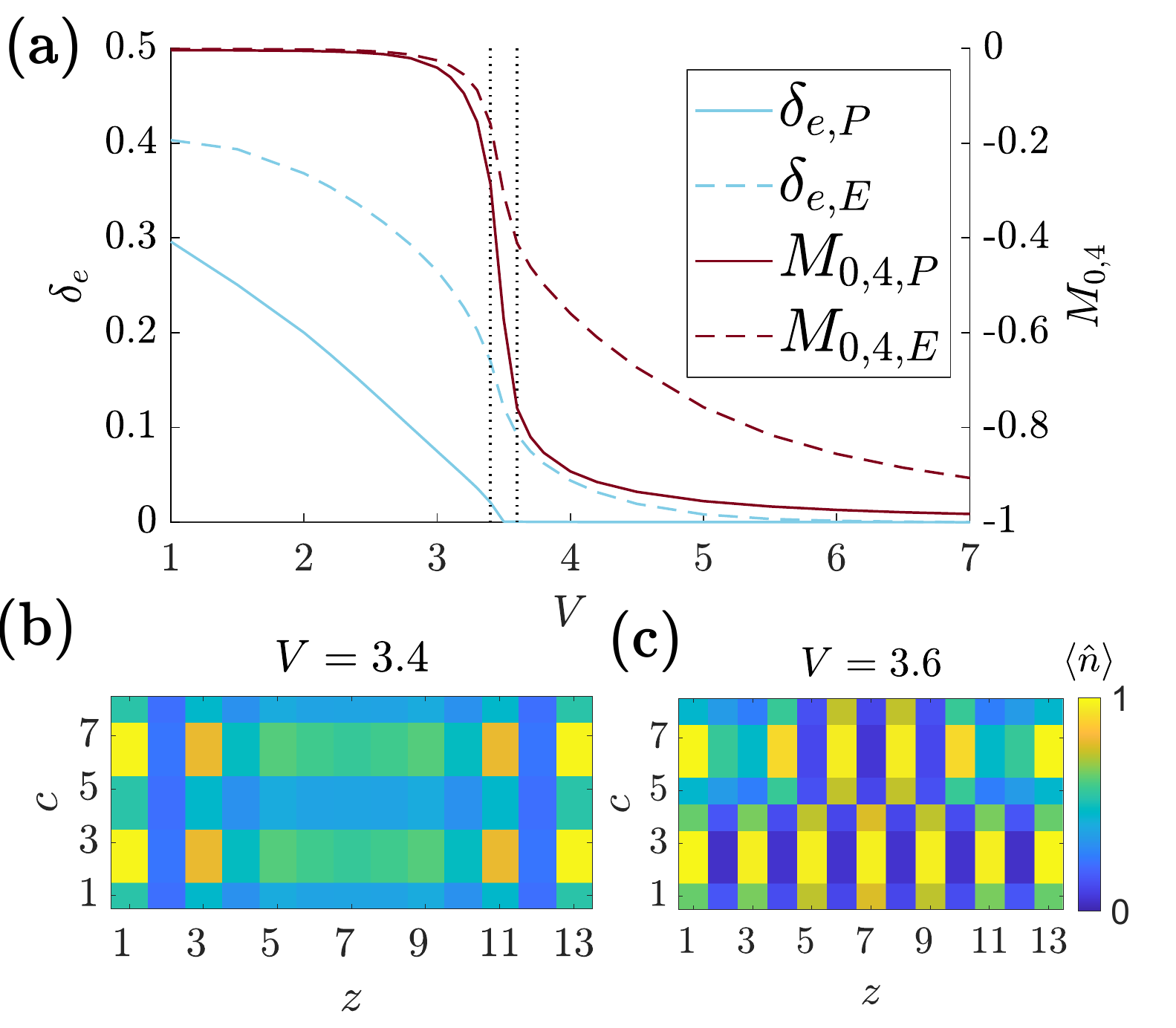}
\caption{Colour online. Further detail for $\theta = \frac{\pi}{2}$. (a) Sublattice-resolved orbital occupation and solid order parameters as a function of $V$. Vertical black dotted lines denote $V = 3.4$ and $V = 3.6$, for which the on-site density is plotted in (b) and (c) respectively. Plots (b) and (c) use the colour scale shown next to (c).}
\label{DCdetailfig}
\end{figure}

At $\theta = \frac{\pi}{2}$, the repulsion between rings creates a `dual checkerboard' (DC) (Figure \ref{phasefig}(e)) in which either $c = $\{$1,2,3,4$\} or $c = $\{$5,6,7,8$\} are occupied on alternate rings. The emergence of the DC as $V$ increases is more complicated than for the other solid states, as highlighted in Figure \ref{DCdetailfig}. A different type of ordering (shown in Figure \ref{DCdetailfig}(b)) emerges from the edges of the cylinder at intermediate values of $V \approx 3$ before the transition to the DC, featuring large occupation of the polar sublattice on alternate rings while the leftover particles are able to tunnel through the other sites. Even for odd system lengths $L_z$ which are the most suited to this partial density ordering, the order does not fully penetrate the bulk of the system before the DC becomes favoured at stronger interaction.

The transition to the DC is much sharper on the polar sublattice than the equatorial sublattice for $L_z = 13$, as shown in Figure \ref{DCdetailfig}(a). The occupation difference of the two dominant modes on the polar sublattice vanishes abruptly as the DC is formed (similar to the TU-SS transition in region $I$), while the suppression of the dominant bosonic orbital is more gradual on the equatorial sublattice (similar to the TU-PS transition in region $II$). The polar sublattice also acquires the DC solid order more sharply than the equatorial sublattice. This example shows how the spatially-varying interactions lead to distinct transition sharpness between the two sublattices in finite-size systems. 

The only other clear qualitative ordering within region $III$ is the diagonal solid (DS) family of states, for which the $L_z = 13$ example is shown in Figure \ref{phasefig}(d). These states emerge as the repulsive axial interactions break apart the PS in the bulk into blocks of four or six occupied sites. The bosons which are moved from the polar sublattice to the equatorial sublattice attach themselves to these blocks in a diagonal pattern which is set by the fact that the next-nearest-neighbour interaction along $\hat{\bm{a}} + \hat{\bm{z}}$ is more attractive than the interaction along $\hat{\bm{a}} - \hat{\bm{z}}$ for $0 < \theta < \frac{\pi}{2}$.

Because uniform states with blocks of four or six occupied polar sites have a similar energy in this parameter range (see appendix section \ref{dssec}), the existence of bulk periodicity (or even a qualitative DS state itself) is very sensitive to edge effects which constrain the total block lengths and the number of repulsive interactions between the blocks. While the shown example ($L_z = 13$) includes both block lengths, our numerical results suggest $L_z = 10$ and $L_z = 16$ favour blocks of four for certain values of $\theta$ which gives a state with a periodicity of three sites in the axial direction. We find that edge effects reduce the stability of this family of states to tunnelling for $L_z = 12$ and $14$, although the qualitative pattern survives for certain values of $\theta$. As with the DC, a higher value of $V$ is required to suppress tunnelling on the equatorial sublattice than the polar sublattice. Although its experimental observation would likely be very challenging due to susceptibility to edge effects, the DS state shows how the sublattice structure, which creates competition between nearest-neighbour interactions when the axial interactions are repulsive, increases the importance of the next-nearest-neighbour interactions to the density ordering. 

We now describe how these two ordered states fit within the RS region. Qualitatively, increasing $\theta$ within region $III$ at strong interaction first splits the PS into progressively smaller blocks which then form DS order within the bulk if allowed by edge effects. For $L_z = 13$, we find the DS exists for $0.72 \frac{\pi}{2} \leq \theta \leq 0.76 \frac{\pi}{2}$. Our numerical results subsequently interpolate from the DS to the DC via a series of states which show domains of both orders which the DC increasingly dominates, although the specific density pattern is highly-sensitive to both edge effects and polarisation direction and is difficult to categorise by solid order. The unifying feature of this region which distinguishes it from the TU is the reduction of condensation and entanglement due to strong density-density interactions. We therefore focus our quantitative analysis to marking a boundary between this repulsive solid (RS) and the TU using local maxima of $S_{ent}$ which are accompanied by decreases in $\delta_e$. We denote this boundary with five-pointed star markers in Figure \ref{phasefig}(a).

It should be noted that the sharpness of this transition is reduced due to the greater tunnelling on the equatorial sublattice as well as the fact that the solid order itself is not always well-defined in this region. This is shown in the $V = 4$ lines in Figure \ref{thetafig}, where the sharp fall in $S_{ent}$ requires a slightly lower value of $\theta$ than the abrupt fall in $M_{0,4}$ which signifies the emergence of DC order. This behaviour contrasts with the TU-SS transition in region $I$ where all physical parameters change together sharply. 
 
\section{Finite-Size Scaling}
\label{ressec_fss}

In this section, we vary $L_z$ to investigate the physics of the TU-SS, TU-PS, and TU-DC transitions in regions I, II, and III respectively. We keep $\theta$ fixed at three separate values and vary $V$ in smaller increments across the transitions.

\subsection{TU-SS Transition}

To study the TU-SS transition in region $I$, we fix $\theta = 0$. To identify the transition point $V_{TU-SS}$, we use the stripe `magnetisation' operator $\hat{m}_{SS} = \frac{2}{L_z L_c} \sum_{z,c} (-1)^c (\hat{n}_{z,c} - \frac{1}{2})$ which takes the values $\pm 1$ in the SS state depending on which of the two degenerate ground states is present. From this we use the Binder cumulant \cite{Binder_cumulant} $U_{SS} = \frac{1}{2}(3 - \frac{\langle \hat{m}_{SS}^4 \rangle}{\langle \hat{m}_{SS}^2 \rangle^2})$, which increases from $0$ to $1$ at the TU-SS transition with increasing sharpness for larger systems. The crossing points of the Binder cumulants as a function of $V$ converge quickly with increasing system size to $V_{TU-SS} \approx 1.46$ which is close to the corresponding transition point found in previous calculations of closely-related models on two-dimensional flat square lattice models \cite{QMCstripe,in_plane_BH_noSS}.

\begin{figure}
\includegraphics[scale=0.45]{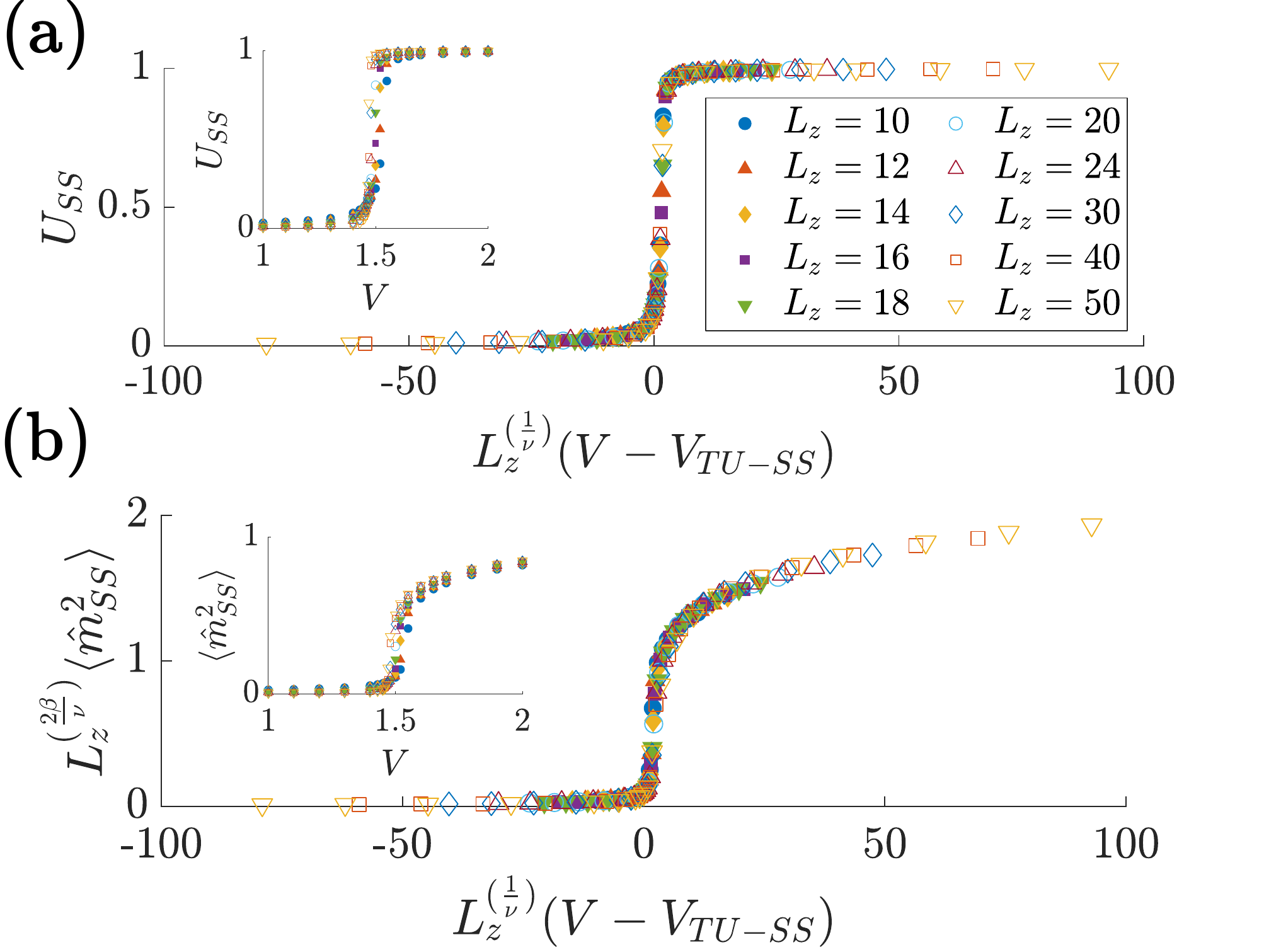}
\caption{Colour online. Finite-size scaling data for TU-SS transition at $\theta = 0$. (a) Binder cumulant $U_{SS}$ against rescaled interaction strength for $\nu = 0.76$. Inset shows $U_{SS}$ against $V$. (b) Rescaled square magnetisation $L_z^{(\frac{2\beta}{\nu})}\langle m_{SS}^2 \rangle$ against rescaled interaction strength for $\beta = 0.08$. Inset shows original square magnetisation $\langle m_{SS}^2 \rangle$ against interaction strength. Markers for different $L_z$ are plotted according to the legend in (a).}
\label{FSS_TUSSfig}
\end{figure}

It is expected that the Binder cumulants for varying $L_z$ will collapse onto each other when plotted against the rescaled interaction strength $L_z^{(\frac{1}{\nu})} (V - V_{TU-SS})$, where $\nu$ is the critical exponent for the correlation length. This is shown in Figure \ref{FSS_TUSSfig}(a) for $\nu = 0.76$. Similar collapse is observed when plotting $L_z^{(\frac{2\beta}{\nu})} \langle m_{SS}^2 \rangle$ against the rescaled interaction strength, where $\beta$ is the critical exponent for the magnetisation, as shown in Figure \ref{FSS_TUSSfig}(b) for $\beta = 0.08$, where we used the measure in Ref. \cite{FSS_quantitative_fit} to quantitatively determine the values of $\nu$ and $\beta$ which produced the best data collapse. The data collapse indicates that the abrupt changes in the behaviour of the small system correspond to a phase transition in the thermodynamic limit.

\subsection{TU-PS Transition}
\label{resssec_TUPS}

\begin{figure}
\includegraphics[scale=0.31]{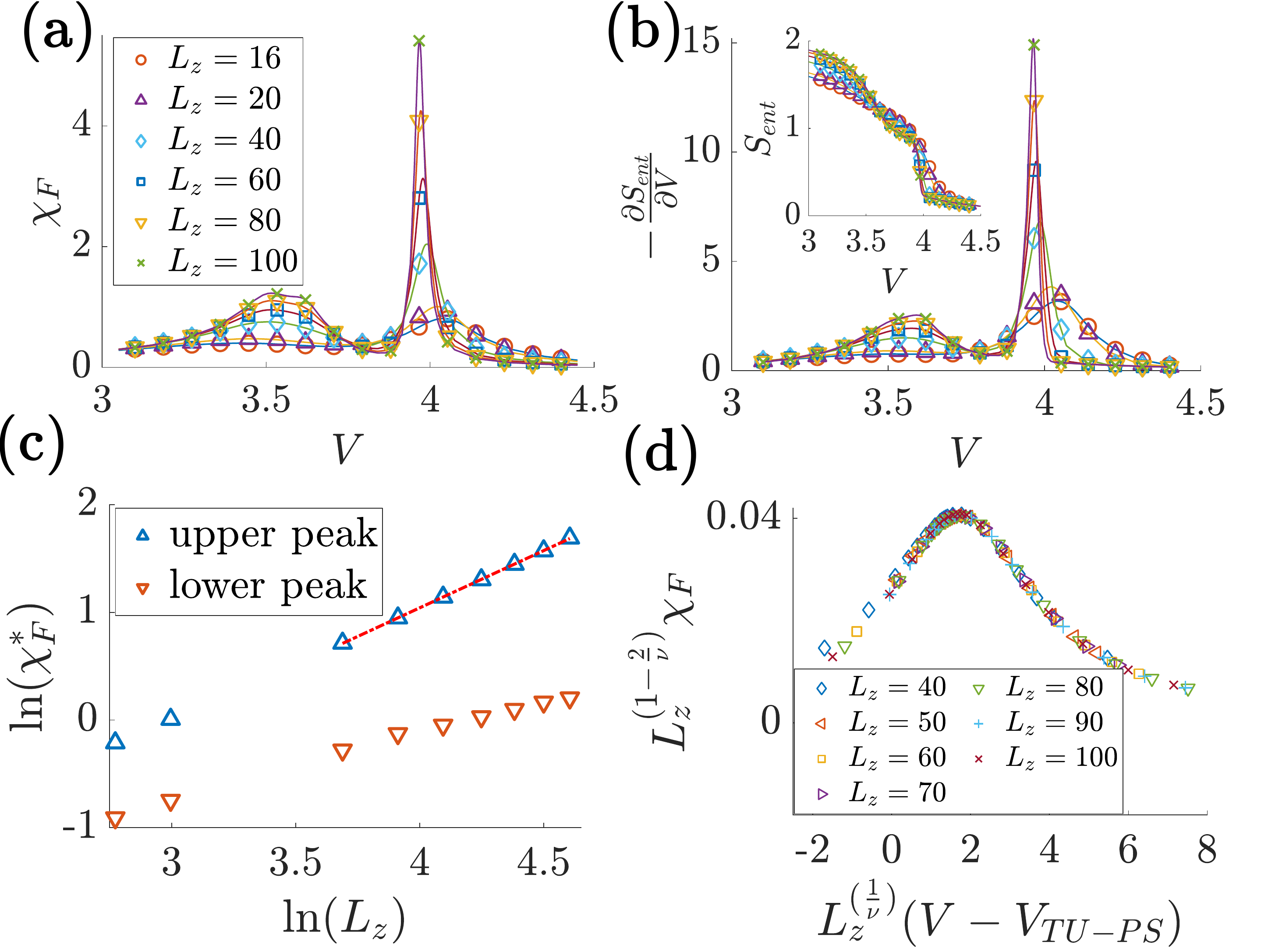}
\caption{Colour online. TU-PS transition at $\theta = \frac{\pi}{4}$ for different $L_z$. (a) $\chi_F$ as a function of $V$. (b) $\frac{\partial S_{ent}}{\partial V}$ for the same values of $L_z$ shown in (a). Inset shows $S_{ent}$ as a function of $V$. (c) $\ln(\chi_F^{*})$ against $\ln(L_z)$ for both peaks. The dashed red line is a linear fit for the upper peak data for the seven largest system sizes $L_z = \{40, 50, 60, 70, 80, 90, 100\}$, from which $\nu = 0.97$ was extracted. (d) Rescaled fidelity susceptibility $L_z^{(1-\frac{2}{\nu})}\chi_F$ against rescaled interaction strength $L_z^{(\frac{1}{\nu})}(V-V_{TU-PS})$ around the upper peak for the seven largest system sizes.} 
\label{FSS_TUPSfig}
\end{figure}

To study the TU-PS transition in region $II$, we fix $\theta = \frac{\pi}{4}$. In section \ref{ressec}, we identified this transition by a peak in $-\frac{\partial S_{ent}}{\partial V}$ and a local minimum in the overlap between neighbouring wavefunctions at $V = 4.1$. To quantify the overlap between neighbouring wavefunctions, we use a finite-difference analogue of the fidelity susceptibility per ring \cite{fidelity_susceptibility},
\begin{equation}
\chi_F = \frac{-2 \ln(\langle \psi(V - \frac{\delta}{2},\theta)| \psi (V + \frac{\delta}{2},\theta)\rangle )}{L_z \delta^2}, \label{fidsusceq}
\end{equation}
where $\delta$ is the difference in $V$ between the Hamiltonians from which the ground-state wavefunctions were calculated. We note that as $L_z$ increases, a second peak in $-\frac{\partial S_{ent}}{\partial V}$ and $\chi_F$ emerges at $V \approx 3.5$ and both peaks become sharper as $L_z$ increases for both measures, as shown in Figure \ref{FSS_TUPSfig}(a) and (b). We note that the total occupation of the polar sublattice increases monotonically with $V$ through both peaks.

To investigate the critical parameters of this transition, we study the scaling of the peak value of $\chi_F$, which we label $\chi_F^*$. We find that $\chi_F^*$ for the upper peak is proportional to $L_z^{\alpha}$ as derived for second-order transitions transitions with correlation length critical exponent $\nu = \frac{2}{\alpha + 1}$ in one dimension \cite{fidelity_susceptibility_QMC, fidelity_susceptibility_critical_exponents}. From this fit for $L_z \geq 40$, we extract $\nu = 0.97$ as shown in Figure \ref{FSS_TUPSfig}(c). We then find good data collapse for $L_z^{(1-\frac{2}{\nu})} \chi_F$ against the rescaled interaction strength $L_z^{(\frac{1}{\nu})} (V - V_{TU-PS})$ as shown in Figure \ref{FSS_TUPSfig}(d), where we use $V_{TU-PS} = 3.953$ to obtain the best fit. We caution that this form of data collapse does not itself guarantee that the upper peak corresponds to a second-order bulk transition, as equivalent scaling collapse has been used to extract $\nu$ for other transition types, such as topological phase transitions \cite{fidelity_susceptibility_topological}.  For the peak at lower $V$, we find $\chi_{F}^*$ is approximately negatively proportional to $\frac{1}{\ln(L_z)}$ for the largest system sizes, which is often observed numerically for Berezinkskii-Kosterlitz-Thouless type transitions \cite{fidelity_BKT, fidelity_Sent_infinite_order}, although significant finite-size corrections mean even longer systems would be required to confirm this relationship.

Physically, the peak at lower $V$ corresponds to a reduction in tunnelling at the centre of the system, as shown by a reduction in both $S_{ent}$ across the centre-most partition and the occupation of the dominant bosonic orbital. for $V$ between the two peaks, the equatorial density shows small oscillations as a function of $z$ decaying slowly into the bulk of the cylinder, which become more noticeable at larger $L_z$. The peak at higher $V$ causes minimal changes in the polar density for all rings except at the edges and coincides with the equatorial density being confined close to the cylinder edges. While the data collapse in Figure \ref{FSS_TUPSfig}(d) signals the values of $\nu$ and $V_{TU-PS}$, a full characterisation of the transition at the upper peak remains open. There are no clear symmetry-breaking order parameters (such as for the TU-SS transition) as would be expected for a second-order bulk transition, while we were unable to extract positive indicators of topologically non-trivial physics from our numerical results. The importance of the occupation of the equatorial sublattice at the cylinder edges in creating the second sharp reduction in $S_{ent}$ (see Appendix \ref{edgesec}) suggests that the transition mechanism behind the upper peak is strongly linked to edge effects.

Overall, we find that the TU-PS transition splits into two distinct peaks in both $-\frac{\partial S_{ent}}{\partial V}$ and $\chi_{F}$ at large $L_z$ with a narrow intermediate region. The upper peak appears to be strongly influenced by the cylinder edges, so it is unclear whether the upper peak or the intermediate region would exist for periodic boundary conditions in the axial direction. We note that while the low entanglement in this region enables the DMRG calculations to reach larger system lengths $L_z \leq 100$, such system lengths may be more experimentally challenging due to inhomogeneity of the lattice potential in the axial direction \cite{cylindrical}

\subsection{TU-DC Transition}
\label{ressec_TUDC}

\begin{figure}
\includegraphics[scale=0.37]{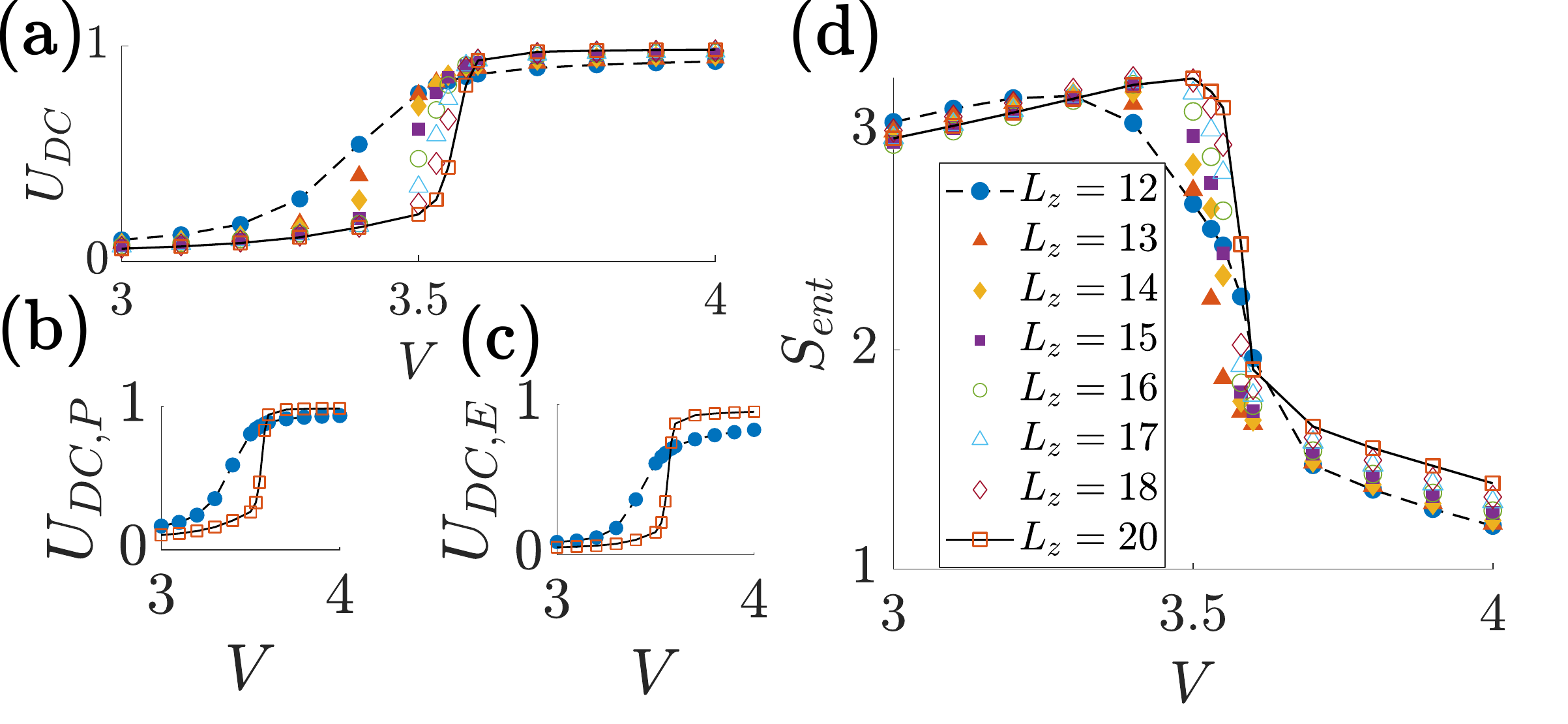}
\caption{Colour online. TU-DC transition at $\theta = \frac{\pi}{2}$ for different $L_z$. (a) Binder cumulant $U_{DC}$ as a function of $V$ for different $L_z$ according to legend in (d). The smallest and largest system sizes ($L_z = 12$ and $L_z = 20$) are plotted with dashed and solid lines respectively between datapoints for clarity. (b) and (c) Sublattice-resolved Binder cumulants for the polar and equatorial sublattices, where only the smallest and largest cylinder lengths are shown for clarity. (d) $S_{ent}$ as a function of $V$.}
\label{FSS_TUDCfig}
\end{figure}

To study the TU-DC transition in region $III$, we fix $\theta = \frac{\pi}{2}$. As the DC state spontaneously breaks a two-fold symmetry, we define the dual-checkerboard magnetisation analagously to the stripe-solid magnetisation, using the formula $\hat{m}_{DC} = \frac{1}{L_z L_c} \sum_{z,c} (-1)^{z+\lfloor \frac{c-1}{4} \rfloor} (\hat{n}_{z,c} - \frac{1}{2})$. We also define polar and equatorial sublattice versions of the dual-checkerboard magnetisation by including only the terms on the relevant sublattice. We then define the Binder cumulant $U_{DC}$ (and its sublattice-resolved versions $U_{DC,P}$ and $U_{DC,E}$) from these magnetisation operators analagously to the TU-SS transition.

We find that $U_{DC}$ increases across the transition more sharply for larger $L_z$ as expected, as is shown in Figure \ref{FSS_TUDCfig}(a), as do its polar and equatorial sublattice counterparts shown in Figure \ref{FSS_TUDCfig}(b) and (c). A similar behaviour is observed for $S_{ent}$ in Figure \ref{FSS_TUDCfig}(d), which peaks more sharply and at slightly larger $V$ for larger $L_z$. This suggests that this transition would become sharper in the thermodynamic limit, however the large entanglement in this region means that confirming this would require larger-scale numerical calculations. 

\section{Discussion}
\label{discsec}

Overall, the octagonal geometry's spatially-dependent interactions lead to richer physics than a flat square lattice or ring lattice, due to the sublattice structure and enhanced frustration. The spatial variation of interactions (which is greater at large $\theta$) favours the occupation of the polar sublattice as predicted previously for a single octagonal ring \cite{ring}, but the interactions between rings can reinforce (in region $II$) or compete with (in region $III$) this tendency depending on the sign of the translationally-invariant axial interactions.

The system demonstrates varied transition behaviours due to the spatially-modulated interactions. In region $I$, where the Hamiltonian is similar to a flat lattice with periodic boundary conditions in one of the two dimensions, the TU-SS transition is sharp even in a small system and we were able to estimate the critical exponents using finite-size scaling. In region $II$, where occupation of the polar sublattice is explicitly favoured rather than spontaneously chosen, the solid-tunnelling transition becomes much more gradual even at large system lengths. Solid order in region $III$, which has the most extreme spatial variation of interactions and has significant occupation of both the polar and equatorial sublattices due to repulsion along the axis, builds sharply on the polar sublattice and gradually on the equatorial sublattice in small systems, although we were unable to confirm whether this sublattice-differentiated behaviour persists in the thermodynamic limit. Spatially-diffuse solid-tunnelling transitions in extended Bose-Hubbard models have previously been predicted due to external mechanisms such as inhomogeneous trapping potentials \cite{inhomogeneous_EBH} and varying coordination number in quasicrystal lattices \cite{quasicrystal_EBH}, but here this effect is driven entirely by interactions.

Importantly for experimental implementation, we find that small axial lengths $L_z$ are generally sufficient to characterise the unusual many-body physics in the bulk of the system and qualitatively identify the phase transitions, which become more pronounced for longer systems. Edge effects are most significant in region $III$ where particles are repelled to the edges, with the DS state being particularly vulnerable to changes in $L_z$, but they also play a role in the TU-PS transition. 

We note that there has been recent interest in simulating spatially-varying lattice Hamiltonians \cite{twistronic_simulation} motivated by the electronic properties of twisted bilayer graphene. This cylindrical model does not offer spatial variation of the Hamiltonian in two lattice dimensions or variation which is incommensurate with the lattice spacing, which prevents Moir\'e supercell structure. However, the general concept of embedding a lower-dimensional system with long-range anisotropic interactions in a higher dimensional space may offer good flexibility to engineer spatially-varying Hamiltonians, particularly for systems built from reconfigurable tweezer arrays \cite{rydberg_arrays_3D}.

\subsection{Additional Parameters}

There are multiple variations of this model which we have not investigated but would offer distinct physics, such as different separations and tunnellings along the axis and around the ring, different filling fractions, removal of the hardcore constraint or dipolar spin models (i.e. particles confined to lattice sites while interacting using the DDI). The value of $L_c$ and polarisation direction in the $x-y$ plane are also important for defining sublattices. For example, a non-trivially different Hamiltonian arises from rotating the polarisation by angles up to $\phi = \frac{\pi}{L_c}$ and $\phi = \frac{\pi}{2L_c}$ for even and odd $L_c$ respectively, which create up to $\frac{L_c}{2}$ or $L_c$ inequivalent classes of site on each ring respectively, compared to the highly-symmetric example studied here which contains two inequivalent classes of site on each ring. 

While our numerical results are limited to small ring lengths $L_c$, it is worth discussing how the interactions change in the (two-dimensional) limit $L_c \rightarrow \infty$ while maintaining the equal separation between neighbouring sites on the ring and axis. In this limit, the discrete ring site index $c$ can be replaced by $\phi_c$ describing the (continuous) azimuthal angle of the site. Unlike the small $L_c$ case we studied, where the interactions vary considerably between neighbouring sites, for infinitely large $L_c$ the interactions between nearby sites are the same as for a flat square lattice with lattice vectors $\hat{\bm{z}}$ and $\hat{\bm{\phi}}$ (while the radial vector $\hat{\bm{R}}$ is perpendicular to the lattice surface). Due to the curved surface, the components of the fixed-space dipole polarisation in these `local' lattice vectors are dependent on the coordinate $\phi_c$ according to the formula.
\begin{equation}
\begin{pmatrix}
d_{\hat{\bm{z}}}\\
d_{\hat{\bm{\phi}}}\\
d_{\hat{\bm{R}}}
\end{pmatrix}
= |d| \begin{pmatrix}
\cos(\theta)\\
-\sin(\theta)\sin(\phi_c)\\
\sin(\theta)\cos(\phi_c)
\end{pmatrix}
\label{curvedipeq}
\end{equation}

This means that the infinite-circumference cylinder implements a continuous series of square-lattice dipolar Bose-Hubbard models joined together, with a slowly-varying effective local polarisation. For example, if $\theta = \frac{\pi}{2}$, the local dipole-dipole interaction at $\phi_c = 0, \pi$ is isotropically repulsive in the lattice plane, which is known to support density waves such as checkerboard and star solids, and supersolid phases \cite{supersolid_square, checker,repBH_ss}. Meanwhile, the same physical polarisation angle $\theta$ causes the polarisation to be along the $\hat{\bm{\phi}}$ lattice vector at $\phi_c = \frac{\pi}{2}, \frac{3 \pi}{2}$, which supports a stripe density wave state \cite{QMCstripe,in_plane_BH_noSS} and a supersolid with the same density order for soft-core bosons \cite{in_plane_soft_core, in_plane_soft_core_2,QMC_stripe_supersolid}. At intermediate $\theta$ and $\phi_c$, the polarisation can point diagonally between $\hat{\bm{z}}$ and $\hat{\bm{\phi}}$ with a perpendicular component in $\hat{\bm{R}}$ providing isotropic repulsion, which has recently been shown to support diagonal stripe and superstripe phases with a $3 \times 3$ unit cell driven by next-nearest-neighbour interactions \cite{phi_BH}. For large $L_c$, the spatially-dependent interactions would enable study of the coexistence and self-organised interfaces between the local states which are supported by given physical values of $\theta$.

\subsection{Physical Implementation}

We have considered the hard-core limit in which the on-site repulsion between two bosons (conventionally labelled $U$ in the Bose-Hubbard model) is sufficiently large that it prevents double occupation of the sites. This limit is useful to avoid complications arising at short distances, such as reactions between polar molecules, but requires strong on-site repulsions. Short-ranged Bose-Hubbard models need $U \gg J$ to prevent tunnelling from favouring a significant probability of double-occupation of a site, while in the extended Bose-Hubbard model, double occupation can also be favoured by off-site interactions. To estimate the necessary value of $U$ to exclude double occupation by this mechanism, we note that the greatest reduction in energy due to double occupancy would occur at $\theta = 0$ by replacing the hard-core stripes by stripes with two particles on each site. Using interactions up to next-nearest-neighbour and ignoring edge effects, the dipole-dipole interaction energy per particle in the hard-core stripe is $-2V$. The double-occupancy stripe would have a larger negative dipole-dipole interaction energy of $-4V$ per particle but would have an extra energy penalty due to the on-site repulsion equalling $\frac{U}{2}$ per particle. We therefore additionally require $U \gg 4V$ in order to prevent the dipole-dipole interaction from favouring double occupation. Alternatively, for weak tunnelling, the existence of reactive losses in polar molecules on the same site can suppress tunnelling to occupied sites \cite{polar_molecules_lattice}.

The numerical calculations presented do not specifically account for finite temperature, but to observe superfluid physics, the tunnelling energy must be greater than the thermal energy. The SS, PS, and DC solid phases require $V/J \geq 2$, $4$, and $4$ respectively implying $V$ must be greater than $4 k_B T$ in order to observe these solids as well. These estimates are broadly in agreement with specific finite temperature calculations for the corresponding flat lattice model at $\theta = 0$ which found a similar bound of $V \gtrapprox 2 k_B T$ for the observation of the SS \cite{QMCstripe}. The energy $V$ for $^{168}$Er magnetic atoms with a lattice spacing of $272$nm is around $h \times 34$ Hz, suggesting temperatures below $0.4$nK would be necessary for $V = 4 k_B T$, compared with temperatures of $\approx 70$nK in previous lattice experiments \cite{EBH_atoms}. The dipole-dipole interaction could be strengthened using Feshbach molecules made from two magnetic atoms \cite{two_magnetic_atoms} or with polar molecules which interact using electric dipole moments. For polar molecules, dipole-dipole interactions $> h \times 1$ kHz are achievable for lattice spacings of $532$nm, meaning $V = 4 k_B T$ corresponds to temperatures below $10$nK, while temperatures of below $60$nK have been recorded in bulk gases of polar molecules \cite{polar_molecules_fermi_degeneracy, low_temperature_bosonic_molecules}. A further consideration for polar molecules however is that the filling fraction of one boson per two lattice sites investigated here is greater than recent experiments where 30$\%$ of sites were occupied \cite{polar_molecules_lattice_4}. While the temperature and filling fraction requirements are challenging for these platforms, we note that experimental progress towards improving these aspects has been rapid and it is plausible that these requirements will be within the reach of near future experiments. Lower temperatures would be needed for the DS, as the next-nearest-neighbour interactions upon which it relies are a factor of $\sqrt{8} \times$ weaker than the nearest-neighbour interactions due to increased distance. 

Another consideration is that the DDI physically acts between all sites whereas our numerical calculations truncated it to next-nearest-neighbour. While shorter-ranged interactions are generally strongest, the long-range tail beyond next-nearest-neighbour can be qualitatively relevant, especially when the shorter-range interactions are significantly frustrated. We note that the curvature of the lattice decreases the physical distance between sites which are separated by $>1$ site along the azimuthal direction, which increases the significance of these long-range interactions. Including the full range of interactions would further stabilise the SS and PS against tunnelling while destabilising the DC, although the weaker interactions at longer range are even more affected by finite temperature.

\section{Conclusions}
We have studied the effect of anisotropic interactions on a real-space cylindrical optical lattice by numerically investigating the zero-temperature ground states of a hard-core dipolar Bose-Hubbard model on a finite-size octagonal-ring cylinder. Compared to flat lattices, the spatially-varying azimuthal density-density interactions offer an additional ordering mechanism which acts either with or against the translationally-invariant axial interactions as controlled by the polarisation direction. We found that this mechanism can directly set the density wave order when cooperating with attractive axial interactions, or accentuate the importance of next-nearest-neighbour interactions and create highly-entangled states and sublattice-differentiated physics when competing with repulsive axial interactions.

\section{Acknowledgements}

The authors would like to acknowledge the use of the University of Oxford Advance Research Computing (ARC) facility in carrying out this work \cite{ARC}. We acknowledge helpful discussions with Joseph Tindall and Paolo Molignini. This work was supported by U.K. Engineering and
Physical Sciences Research Council (EPSRC) Grant
EP/P01058X/1.

\bibliography{planbib} 

\appendix
\beginsupplement
\section{Calculation Details}

\label{calcsec}

To map the 2D cylindrical lattice to a 1D matrix product state, we used a `coil' mapping in which the MPS loops around each ring successively from $c = 1$ to $c = 8$. This means that neighbouring sites along the axis of the cylinder are separated by $L_c$ sites in the 1D chain, which dramatically increases the range the interactions in the 1D chain compared to the 2D `physical' lattice. The 2D-to-1D mapping constrains the possible bipartitions of the MPS that can be used to calculate $S_{ent}$ because the MPS, not only the 2D lattice, must be split into exactly two parts. For example, this mapping prevents partitioning the lattice along the cylinder axis. Mappings which would allow partitions along the axis would also significantly increase computational cost due to the physical periodic boundary conditions in the azimuthal direction.

We use $\approx 80$ DMRG sweeps, of which the majority are performed at low values of $\chi \leq 400$, to approximate the ground state of each different Hamiltonian. We increase $\chi$ until the maximum truncation errors in the DMRG sweep reduce to $\approx 10^{-6}$ and the energy change upon increasing $\chi$ by over $10\%$ is $\approx 1$ part in $10^5$. While this required $\chi = 6400$ for the calculations with the greatest entanglement, $\chi \leq 2000$ was sufficient to ensure acceptable convergence for regions $I$ and $II$. Our calculations use the $U(1)$ symmetry corresponding to conservation of particle number. We use a noise term, which is gradually reduced and then removed in later sweeps, to encourage correlations between sites which were well-separated in the 1D chain \cite{noise_term}.

\section{SS-PS Boundary}

\label{sspssec}

Assuming $V >> J$, the SS-PS transition can be found by calculating the interaction energy of each state as a function of $\theta$. When considering up to next-nearest-neighbour interactions, the SS only features interaction along the $\hat{\bm{z}}$ direction. The interaction energy between two particles occupying adjacent sites along the axis is $(1-3\cos^2(\theta))V$. For a finite octagonal prism lattice of length $L_z$, there are $4(L_z - 1)$ such interactions.

The PS has the same interactions along $\hat{\bm{z}}$, but has additional interactions along $\hat{\bm{x}}$ and next-nearest-neighbour interactions along $\hat{\bm{x}} \pm \hat{\bm{z}}$. The energy for each interaction along $\hat{\bm{x}}$ is $(1-3\sin^2(\theta))V$ and there are $2L_z$ such interactions. The angular dependence of the interactions along $\hat{\bm{x}} + \hat{\bm{z}}$ and $\hat{\bm{x}} - \hat{\bm{z}}$ cancel when added together. The sum of these two interactions is $-\frac{1}{\sqrt{8}}V$ and there are $2(L_z - 1)$ such pairs of interactions. This means that the SS and PS states have equal energy when $L_z (1-3\sin^2(\theta)) = \frac{L_z - 1}{\sqrt{8}}$.

\section{Variations of DS state}

\label{dssec}

\begin{figure}
\includegraphics[scale=0.33]{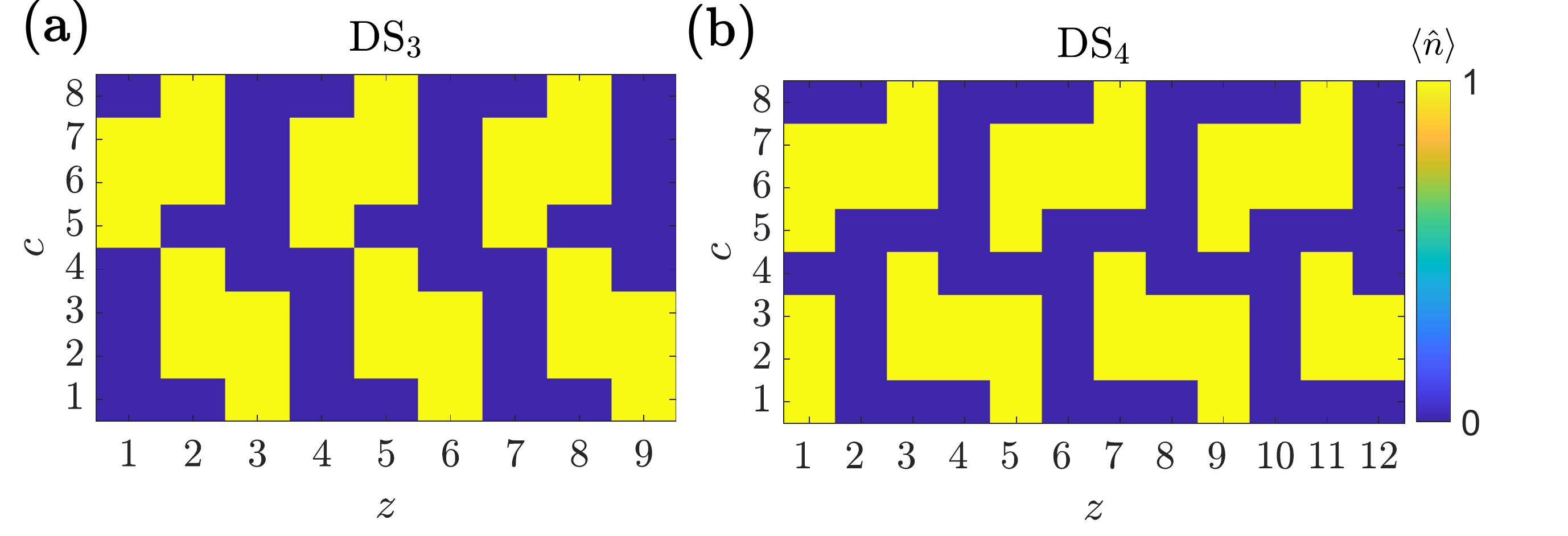}
\caption{Colour online. $\langle \hat{n}_{z,c} \rangle$ for three periods of (a) DS$_3$ (b) DS$_4$ states.}
\label{DSfig}
\end{figure}

In this section, we focus on the limit of strong DDI and discuss the role of edge effects in stabilising variations of the DS state with different block sizes. We label these states by their periodicity, where Figure \ref{DSfig} shows the on-site density for the DS$_3$ and DS$_4$ variations as examples. Note that this family of states also includes the DC, which can be labelled as DS$_2$, and the PS, which can be labelled as DS$_{\infty}$. The sublattice ordering induced by the spatially-dependent azimuthal interactions favours higher values of the periodicity to increase polar sublattice occupation while repulsion along the cylinder axis favours lower values of periodicity.

With interactions truncated to next-nearest-neighbour, the highest periodicity which can be favoured is the DS$_4$ state. Relative to the DS$_3$ variation, this state benefits from the removal of one repulsive interaction in the $\hat{\bm{y}} + \hat{\bm{z}}$ direction per period. This means that the DS$_4$ state has a lower energy than the other states in the DS family for $0.67 \frac{\pi}{2} \leq \theta \leq 0.75 \frac{\pi}{2}$. However, edge effects nullify the advantage of avoiding the repulsive $\hat{\bm{y}} + \hat{\bm{z}}$ interaction and favour DS$_3$ ordering close to the edges. For $L_z = 13$ this results in the DS taking the form shown in Figure \ref{phasefig}(d), where the domains of DS$_3$ are close to the edges and the domains of DS$_4$ are in the centre of the cylinder. 

We note that this pattern can be complicated by the fact that the particle blocks must be compatible with the cylinder length, which also often results in blocks of two polar sublattice particles at the edges as shown in Figure \ref{phasefig}(c). By calculating the interaction energy of all DS-type product states (including the possibility of these edge blocks) for $10 \leq L_z \leq 16$ with one boson per two sites, we find that the energy landscape is a delicate function of how the blocks fit with the edges and whether this allows the DS$_4$ blocks to reduce repulsion between blocks as it would for a periodic DS$_4$ state. This creates a large number of product states which are close in energy, which increases the difficulty in numerically determining the ground state for finite tunnelling. By comparing DMRG results for initial states set either to random states with small $\chi$ or the DS-type product state with the lowest interaction energy, we find that our numerical results for all studied values of $L_z$ feature a qualitative DS-type ground state for certain values of $\theta$ in the range $0.72 \frac{\pi}{2} \leq \theta \leq 0.77 \frac{\pi}{2}$ at the strong interaction strength of $V=7$ but for other values of $\theta$ within this range even the initialised DS-type product states can become significantly distorted during optimisation, especially for $L_z = 12$ and $16$. 

\section{Varying Particle Number for TU-PS Transition}
\label{edgesec}

\begin{figure}
\includegraphics[scale=0.35]{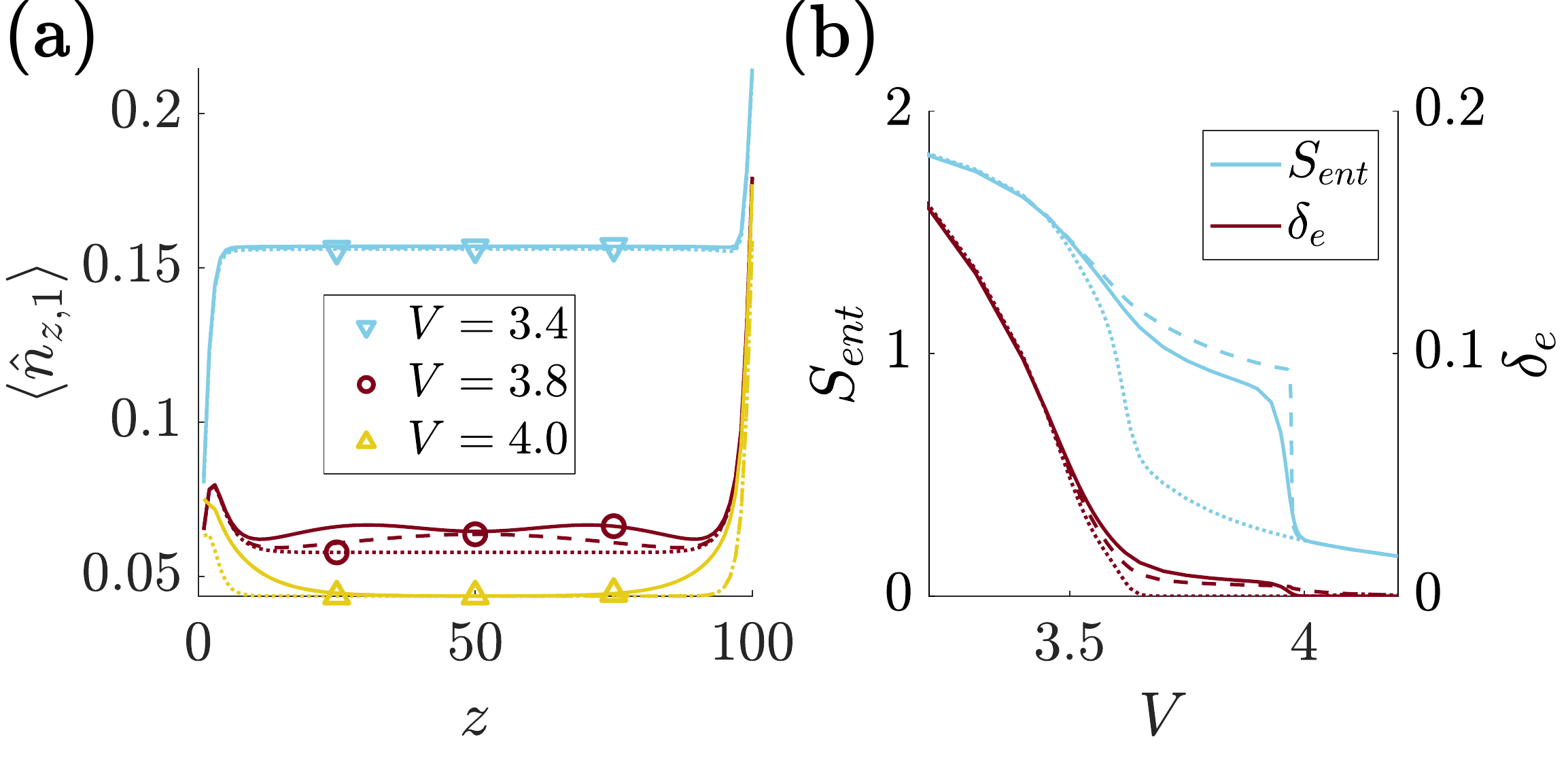}
\caption{Colour online. TU-PS transition at $\theta = \frac{\pi}{4}$ for $L_z = 100$ plotted with $N = N_H$ in solid lines, $N = N_H - 1$ in dashed lines, and $N = N_H - 2$ in dotted lines. (a) $\langle \hat{n}_{z,c = 1} \rangle$ for three different values of $V$. Neighbouring densities for the discrete site index $z$ are joined by straight lines to guide the eye, while one marker per line is used for identification. (b) $S_{ent}$ and $\delta_e$ as a function of $V$.}
\label{TUPS_varyNfig}
\end{figure}

To show the role of the cylinder edges in the TU-PS transition, we compare results for $\theta = \frac{\pi}{4}$ and $L_z = 100$ for three fixed particle numbers: $N = N_H$ (the same filling fraction used for all other results in this paper), $N = N_H - 1$, and $N = N_H - 2$. We find that removing one or two particles has a large impact on the density at the cylinder edges and on the features of $S_{ent}$ which physically signify the upper part of the TU-PS transition.

In Figure \ref{TUPS_varyNfig}(a) we show the on-site density for $c = 1$ on the equatorial sublattice for three qualitatively different interaction strengths: $V = 3.4$ (below both peaks in $-\frac{\partial S_{ent}}{\partial V}$), $V = 3.8$ (between the two peaks in $-\frac{\partial S_{ent}}{\partial V}$), and $V = 4.0$ (above both peaks in $-\frac{\partial S_{ent}}{\partial V}$). We firstly note that the density at the two edges is different because for $c = 1,8$, the cylinder edge at $z = 1$ is less occupied than the edge at $z = L_z$, while the opposite is true for $c = 4,5$. For $V = 3.4$, the equatorial density is uniform in the bulk, while for $V = 3.8$ it oscillates in the bulk, and for $V = 4.0$ it is confined to the edges. The change in density upon removing particles is also qualitatively different for the chosen values of $V$: For $V = 3.4$, the change in density is distributed uniformly across the bulk. For $V = 3.8$, adding one and then two particles to $N = N_H - 2$ adds one and two peaks respectively to the equatorial density distribution. For $V = 4.0$, the change in equatorial density upon removing particles is confined to the edges and the equatorial density distribution for $N = N_H - 1$ overlaps with that of $N = N_H$ at one edge and $N = N_H - 2$ at the other edge. (We have confirmed that for $N = N_H - 1$ there are two degenerate states at which the edge particle/hole is swapped to the other edge.)

In Figure \ref{TUPS_varyNfig}(b) we show $S_{ent}$ and $\delta_e$ for the same three particle numbers. For $3.5 \leq V \leq 4.0$, $S_{ent}$ depends strongly on the exact number of particles and reduces sharply when the density on the equatorial sublattice becomes confined to the edges. A very small non-zero value of $\delta_e$ also survives in this region for $N = N_H, N_H - 1$. The reduction is $S_{ent}$ is particularly sharp for $N = N_H - 1$, where the choice of which edge of the cylinder has the particle/hole spontaneously breaks a two-fold degeneracy at $V \approx 4$. These observations suggest that the part of the TU-PS transition at higher $V$ in the finite cylinder is strongly affected by the cylinder edges.

\end{document}